\documentclass[preprint,nofootinbib,aps,superscriptaddress]{revtex4}

\newcommand{\beq}{\begin{equation}}
\newcommand{\eeq}{\end{equation}}
\newcommand{\beqa}{\begin{eqnarray}}
\newcommand{\eeqa}{\end{eqnarray}}
\newcommand{\no}{\nonumber}

\begin{document}

\preprint{\vbox{\hbox{LBNL-52330} \hbox{WIS/2/03-Mar-DPP}
  \hbox{SLAC-PUB-9670} \hbox{hep-ph/0303171}}}

\vspace*{1cm}

\title{\boldmath SU(3) Relations and the CP Asymmetries in $B$ Decays\\
  to $\eta^\prime K_S$, $\phi K_S$ and $K^+K^-K_S$}

\author{Yuval Grossman}\email{yuvalg@physics.technion.ac.il}
\affiliation{Department of Physics, Technion--Israel Institute of
  Technology\\ Technion City, 32000 Haifa, Israel}

\author{Zoltan Ligeti}\email{zligeti@lbl.gov}
\affiliation{Ernest Orlando Lawrence Berkeley National Laboratory \\
  University of California, Berkeley, CA 94720, USA}

\author{Yosef Nir}\email{yosef.nir@weizmann.ac.il}
\affiliation{Department of Particle Physics \\
  Weizmann Institute of Science, Rehovot 76100, Israel}

\author{Helen Quinn}\email{quinn@slac.stanford.edu}
\affiliation{Stanford Linear Accelerator Center\\
  Stanford University, Menlo Park, CA 94025, USA\\[-12pt] $\phantom{}$}

%\date{\today}
%\pacs{12.10.Dm, 12.10.Kt, 98.80.Cq}

\begin{abstract}

We consider CP asymmetries in neutral $B$ meson decays to $\eta' K_S$,
$\phi K_S$, and $K^+K^-K_S$. We use SU(3) relations to estimate or
bound the contributions to these amplitudes proportional to
$V_{ub}^*V_{us}$. Such contributions induce a deviation of the $S_f$
terms measured in these time dependent CP asymmetries from that
measured for $\psi K_S$. For the $K^+K^-K_S$ mode, we estimate the
deviation to be of order 0.1. For the $\eta^\prime K_S$ mode, we
obtain an upper bound on this deviation of order 0.35. For the $\phi
K_S$ mode, we have to add a mild dynamical assumption to the SU(3)
analysis due to insufficient available data, yielding an upper bound
of order 0.25. These bounds may improve significantly with future
data.  While they are large at present compared to the usually assumed
Standard Model contribution, they are obtained with minimal
assumptions and hence provide more rigorous tests for new physics.  If
measurements yield $|S_f-S_{\psi K}|$ that are much larger than our
bounds, it would make a convincing case for new physics.

\end{abstract}

\maketitle

%%%%%%%%%%%%%%%%%%%%%%%%%%%%%%%%%%%%%%%%%%%%%%%%%%%%%%%%%%%%%%%%%%%%%%
\section{Introduction}
\label{sec:introduction}

Recent measurements of CP asymmetries in neutral $B$ meson decays into
final CP eigenstates test the Kobayashi-Maskawa mechanism and probe
new sources of CP violation. The time dependent asymmetries depend on
two parameters, $S_f$ and $C_f$ ($f$ denotes here a final CP eigenstate):
\beq\label{defcs}
{\cal A}_f(t)\equiv\frac
{\Gamma(\bar B^0_{\rm phys}(t)\to f)-\Gamma(B^0_{\rm phys}(t)\to f)}
{\Gamma(\bar B^0_{\rm phys}(t)\to f)+\Gamma(B^0_{\rm phys}(t)\to f)}
=-C_f\cos(\Delta m_B\, t)+S_f\sin(\Delta m_B\,t)\,.
\eeq
CP violation in decay induces $C_f$, while CP violation in the interference of
decays with and without mixing induces $S_f$. (The contribution from CP
violation in mixing is at or below the percent level and can be safely
neglected with the present experimental accuracy.)

If the decay is dominated by a single weak phase, $C_f\approx0$ and the value
of $S_f$ can be cleanly interpreted in terms of CP violating parameters of the
Lagrangian. This is the case for decays which are dominated by the tree $b\to
c\bar cs$ transition or by the gluonic penguin $b\to s\bar ss$ transition. If
one neglects the subdominant amplitudes with a different weak phase, the CP
asymmetries in these two classes of decays are given by
$S_f=-\eta_f\sin2\beta$, where $\eta_f=+1(-1)$ for final CP-even (-odd) states
and $\beta$ is one of the angles of the unitarity triangle. In particular, in
this approximation, the CP asymmetries in the two classes are equal to each
other, for example, $S_{\psi K}=S_{\phi K}$.  A strong violation of such a
relation would indicate new physics~\cite{Grossman:1996ke}.

Our aim in this paper is to quantify this statement with minimal assumptions
for three modes of interest: $\phi K_S$, $\eta^\prime K_S$ and $K^+K^-K_S$. We
would like to estimate or to find bounds on the deviations of the
corresponding  asymmetries from $S_{\psi K_S}$ that are (hadronic-)model
independent. The ingredients of our analysis are SU(3) relations and
experimental information on related modes. We will be able to carry out this
program to the end for $S_{\eta^\prime K_S}$. As concerns $S_{\phi K_S}$, we
derive SU(3) relations that can, in principle, lead to model independent
bounds. In practice, however, some experimental information is still missing.
Nevertheless, by using a mild dynamical assumption, we obtain a bound for this
mode too. The situation is more complicated for $S_{KKK}$, where we point out
some subtleties in the interpretation of the experimental results. For this
mode, however, we are able to estimate (rather than just bound) the deviation
of the extracted asymmetry from $\sin2\beta$ in the Standard Model by using
U-spin relations and  experimental data.

%%%%%%%%%%%%%%%%%%%%%%%%%%%%%%%%%%%%%%%%%%%%%%%%%%%%%%%%%%%%%%%%%%%%%%
\section{Experimental and Theoretical Background}
\label{sec:data}

The CP asymmetry in $B\to\psi K_S$ decays (and other, related, modes
that proceed via $b\to c\bar cs$) has been measured, with a world
average \cite{Nir:2002gu} of
\beqa\label{psikwa}
S_{\psi K_S}&=&+0.734\pm0.054\quad \mbox{\cite{Aubert:2002ic,Abe:2002px}},\no\\
C_{\psi K_S}&=&+0.05\pm0.04\qquad \mbox{\cite{Aubert:2002ic,Abe:2002px}}.
\eeqa
The value of $S_{\psi K_S}$ is consistent with predictions made on the
basis of other measurements of the CKM parameters ($\Delta m_B$,
$\Delta m_{B_s}$, $\varepsilon_K$ and tree level decays).

CP asymmetries have also been searched for in three modes that are
dominated by $b\to s\bar ss$ gluonic penguin transitions:
\beqa\label{etakwa}
S_{\eta^\prime K_S}&=&+0.33\pm0.34\quad
  \mbox{\cite{Abe:2002np,Aubert:2003bq}},\no\\
C_{\eta^\prime K_S}&=&-0.08\pm0.18\quad
  \mbox{\cite{Abe:2002np,Aubert:2003bq}},\\[4pt]
\label{phikwa}
S_{\phi K_S}&=&-0.39\pm0.41\quad
  \mbox{\cite{Abe:2002np,Aubert:2002nx}},\no\\
C_{\phi K_S}&=&+0.56\pm0.44\quad
  \mbox{\cite{Abe:2002np}},\\[4pt]
\label{kkkwa}
-S_{K^+K^- K_S}&=&+0.49\pm0.44^{+0.33}_{-0.00}\quad
  \mbox{\cite{Abe:2002np}},\no\\
C_{K^+K^- K_S}&=&+0.40\pm0.34^{+0.26}_{-0.00}\quad
  \mbox{\cite{Abe:2002np}}.
\eeqa
The Standard Model predicts that in these modes $-\eta_f S_f=S_{\psi K_S}$
and $C_f=0$ to a good approximation. The statistical errors in
Eqs.~(\ref{etakwa})--(\ref{kkkwa}) are too large to make any firm conclusions.
It is clear, however, that there is still much room left for deviations from
the Standard Model generated by possible new physics in $b\to s$ transitions.

The Standard Model amplitude for these three decay modes can be
written as follows:
\beq\label{abqqs}
A_f \equiv A(B^0\to f) = V_{cb}^*V_{cs}\ a_f^c+V_{ub}^*V_{us}\ a_f^u\,.
\eeq
The second term is CKM-suppressed compared to the first one since
\beq\label{ckmsup}
{\cal I}m\! \left({V_{ub}^*V_{us}\over V_{cb}^*V_{cs}}\right)
= \left|{V_{ub}^*V_{us}\over V_{cb}^*V_{cs}}\right| \sin\gamma
= {\cal O}(\lambda^2)\,,
\eeq
where $\lambda=0.22$ is the Wolfenstein parameter.
It is convenient to define
\beq\label{defxi}
\xi_f \equiv \frac{V_{ub}^*V_{us}\ a_f^u}{V_{cb}^*V_{cs}\ a_f^c}\,,
\eeq
and thus rewrite the amplitude of Eq.~(\ref{abqqs}),
\beq
A_f =V_{cb}^*V_{cs}\, a_f^c\, (1+\xi_f)\,.
\eeq
The SU(3) analysis we carry out allows us to bound $|\xi_f|$. To first order in
this quantity, the deviation of the asymmetry from $\sin2\beta$ is given by
\cite{Gronau:1989ia,Grossman:2002bu}
\beq\label{sminusstb}
-\eta_f S_f-\sin2\beta = 2\cos2\beta\, \sin\gamma\, \cos\delta_f\; |\xi_f|\,,
\eeq
where $\delta_f=\arg(a^u_f/a^c_f)$. The $\xi_f$ parameter characterizes
also the size of $C_f$:
\beq
C_f = -2\sin\gamma\, \sin\delta_f\; |\xi_f|\,,
\eeq
Note that $\delta_f$ can be determined from
$\tan\delta_f=(\eta_fS_f+\sin2\beta)/(C_f \cos2\beta)$, while
the following ($\delta_f$-independent) relation between $S_f$, $C_f$ and
$|\xi_f|$ may become useful in the future:
\beq\label{cfsf}
C_f^2+[(\eta_fS_f+\sin2\beta)/\cos2\beta]^2=4\sin^2\gamma\, |\xi_f|^2.
\eeq

The crucial question, when thinking of the deviation of $-\eta_f S_f$ from
$\sin2\beta$, is the size of $a_f^u/a_f^c$.  While $a_f^c$ is dominated by the
contribution of $b \to s\bar ss$ gluonic penguin diagrams, $a_f^u$ gets
contributions from both penguin diagrams and $b \to u\bar u s $ tree diagrams.
For the penguin contributions, it is clear that $|a_f^u/a_f^c| \sim 1$. (The
$a_f^c$ term comes from the charm penguin minus the top penguin, while the up
penguin minus the top penguin contributes to $a_f^u$.) Thus our main concern is
the possibility that the tree contributions might yield $|a_f^u/a_f^c|$
significantly larger than one.

For final states with zero strangeness, $f'$, we write the amplitudes as
\beq\label{abuud}
A_{f^\prime} \equiv A(B^0\to f') = V_{cb}^*V_{cd}\ b_{f^\prime}^c
  + V_{ub}^*V_{ud}\ b_{f^\prime}^u\,.
\eeq
Here neither term is CKM suppressed compared to the other. We use SU(3) flavor
symmetry to relate the $a^{u,c}_f$ amplitudes to sums of $b_{f^\prime}^{u,c}$.
While similar SU(3) relationships have been explored elsewhere~\cite{Savage:ub,
Grinstein:1996us, Grossman:1997gr, Dighe:1997wj, Deshpande:2000jp}, most of our
results and applications are new.

The SU(3) relations, together with the measurements or upper bounds on the
rates for the non-strange channels plus the measured rate for the channel of
interest yield an upper bound on $|\xi_f|$.  Let us first provide an  intuitive
explanation of this.  The decays to final strange states, $f$, are dominated by
the $a_f^c$ terms.  Those to final states with zero strangeness, $f'$, are
dominated by the $b_{f^\prime}^u$ terms. Thus we can estimate  $|a_f^c|$ and
$|b_{f^\prime}^u|$ from the measured branching ratios (or the upper bounds on
them).  Then the SU(3) relations give upper bounds on certain sums of the
$b_{f^\prime}^c$ and $a_f^u$ amplitudes from the extracted values of $a_f^c$
and $b_{f^\prime}^u$, respectively. This then gives a bound on $|a_f^u/a_f^c|$,
and consequently on $|\xi_f|$.  We can also check the self-consistency of the
analysis, namely that $|a_f^u| < |A_f|/|V_{ub}V_{us}|$ and $|b_{f^\prime}^c| <
|A_{f^\prime}|/|V_{cb}V_{cd}|$.  However, as we show below, the assumptions
made in this paragraph can be avoided entirely.

The SU(3) relations actually provide an upper bound on $|V_{cb}^*V_{cd}\, a_f^c
+ V_{ub}^*V_{ud}\, a_f^u|$, in terms of the measured branching ratios of some
zero strangeness final states (or limits on them).  Therefore, without any
approximations, we can bound
\beq\label{xiratio}
\widehat \xi_f \equiv
\left|\frac{V_{us}}{V_{ud}} \times \frac{V_{cb}^*V_{cd}\ a_f^c
  + V_{ub}^*V_{ud}\ a_f^u}{V_{cb}^*V_{cs}\ a_f^c + V_{ub}^*V_{us}\ a_f^u}\right|
= \left|\frac{\xi_f + (V_{us}V_{cd})/(V_{ud}V_{cs})}{1+\xi_f}\right|.
\eeq
If the bound on $\widehat \xi_f$ is less than unity, then it gives a bound on
$|\xi_f|$.  We work to first order in $\xi_f$, since the naive expectation is
$\xi_f = {\cal O}(\lambda^2)$.  At the present state of the data the bounds we
obtain on $\widehat \xi_f$ are significantly larger that $\lambda^2$, so we
also work in the approximation $\lambda^2 \ll \widehat \xi_f < 1$.  This is
appropriate because we want to constrain the possibility $|a_f^u/a_f^c| \gg
1$.  Therefore, we take $|\xi_f| \approx \widehat\xi_f$ in what follows
(although this approximation should not be made once the bounds on $\widehat
\xi_f$ are of order $\lambda^2$).

The SU(3) decomposition of $a_f^u$ and $b_{f'}^u$ is identical with
that of $a_f^c$ and $b_{f'}^c$ although the values of the reduced
matrix elements are independent for the $u$- and the $c$-terms. The
SU(3) decomposition is given in Appendix~\ref{sec:suthree} for the
channels discussed in this paper. We use the notation $a(f) \equiv
a_f^{u,c}$ and $b(f^\prime) \equiv b_{f'}^{u,c}$ for equations that
apply for both cases, with either all $u$ or all $c$ upper indices.
Our normalization of the various amplitudes is the same as that of
Ref.~\cite{Grinstein:1996us}. It corresponds to $\Gamma=|A|^2$
independent of whether the final particles are identical or not.

The contributions to $a^c_f$ and $b_{f'}^c$ come from penguin
diagrams or the tree $b\to c \bar c q$ transition plus some form of
rescattering (such as $D$-exchange) to replace the $c \bar c $ with lighter
quark flavors. Aside from small electroweak penguin contributions, there is
only an SU(3) triplet term in the Hamiltonian for these amplitudes.  Neglecting
electroweak penguins would result in additional SU(3) relations between the
$a_f^c$ and $b_{f^\prime}^c$ terms.  We do not make such an approximation in
our analysis, but it might be useful for other purposes.

%%%%%%%%%%%%%%%%%%%%%%%%%%%%%%%%%%%%%%%%%%%%%%%%%%%%%%%%%%%%%%%%%%%%%%
\section{\boldmath The CP Asymmetry in $B\to\eta^\prime K_S$}
\label{sec:etak}
%%%%%%%%%%%%%%%
%%%%%%%%%%%%%%%
\subsection{SU(3) relations}
\label{subsec:etapsut}

The CP asymmetry in $B\to\eta^\prime K_S$ is expected to yield a less accurate
measurement of $\sin2\beta$ than the $\psi K_S$ mode. The reason is that, while
this decay is dominated by gluonic penguins, there are CKM-suppressed tree
contributions that induce a deviation from the leading result. Nevertheless, it
was argued in Ref.~\cite{London:1997zk} that this deviation is below the two
percent level. The argument was based on relating the tree contributions in
$B\to\eta^\prime K$ and $B\to\pi\pi$ decays. While this may be a reasonable
hypothesis, it is based on neither approximate symmetry nor obvious dynamical
assumptions. In this section we derive a more rigorous (though weaker) bound on
the ``problematic" subleading contribution.

The results of Appendix~\ref{sec:suthree} imply the following amplitude
relations:
\beqa\label{etaoe}
a(\eta_1 K^0)&=&-\frac{1}{\sqrt2}\, b(\eta_1\pi^0)
  +\sqrt{\frac{3}{2}}\, b(\eta_1\eta_8)\,,\no\\
a(\eta_8 K^0)&=&\frac{1}{2\sqrt2}\, b(\eta_8\pi^0)
  - {\sqrt 3\over 4} \left[b(\pi^0\pi^0)-b(\eta_8\eta_8)\right].
\eeqa
The states $\eta_1$ and $\eta_8$ transform as a singlet and an octet of SU(3),
respectively. They are related to the physical $\eta^\prime$ and $\eta$ states
through an orthogonal rotation:
\beq\label{etaetap}
\eta^\prime=c\,\eta_1-s\,\eta_8, \qquad \eta=s\,\eta_1+c\,\eta_8,
\eeq
where $s\equiv\sin\theta_{\eta\eta^\prime}$ and
$c\equiv\cos\theta_{\eta\eta^\prime}$.  Most extractions of the mixing angle,
$\theta_{\eta\eta^\prime}$, vary in the $10-20^\circ$ range, and we will use
$\theta_{\eta\eta^\prime} = 20^\circ$ in our numerical
calculations~\cite{Hagiwara:fs}.

In terms of physical states, we obtain from Eq.~(\ref{etaoe}) the relation
\beqa\label{etapks}
a(\eta^\prime K^0)&=&
\frac{s^2-2c^2}{2\sqrt2}b(\eta^\prime\pi^0)
-\frac{3sc}{2\sqrt2}b(\eta\pi^0)
+\frac{\sqrt3s}{4}b(\pi^0\pi^0)\no\\
&-&\frac{\sqrt3s(s^2+4c^2)}{4}b(\eta^\prime\eta^\prime)
+\frac{3\sqrt3sc^2}{4}b(\eta\eta)
+\frac{\sqrt3c(2c^2-s^2)}{2\sqrt2}b(\eta\eta^\prime).
\eeqa
The SU(3) analysis gives many more relations, involving both charged and
neutral $B$ decay amplitudes. The most general such relation, involving up to
thirteen amplitudes on the right hand side, is given in
Appendix~\ref{sec:genetapri}. With current data, Eq.~(\ref{etapks}) gives the
strongest bound:
\beqa\label{boundptetap}
|\xi_{\eta^\prime K_S}| &<& \left|\frac{V_{us}}{V_{ud}}\right|
\left[ 0.59\sqrt{\frac{{\cal B}(\eta^\prime\pi^0)}{{\cal B}(\eta^\prime K^0)}}
+0.33\sqrt{\frac{{\cal B}(\eta\pi^0)}{{\cal B}(\eta^\prime K^0)}}
+0.14\sqrt{\frac{{\cal B}(\pi^0\pi^0)}{{\cal B}(\eta^\prime K^0)}}
\right.\no\\
&&+\left. 0.53 \sqrt{\frac{{\cal B}(\eta^\prime\eta^\prime)}{{\cal
B}(\eta^\prime K^0)}} +0.38\sqrt{\frac{{\cal B}(\eta\eta)}{{\cal
B}(\eta^\prime K^0)}} +0.96\sqrt{\frac{{\cal
B}(\eta\eta^\prime)}{{\cal B}(\eta^\prime K^0)}}\right].
\eeqa
This bound is obtained from Eq.~(\ref{etapks}) by taking all amplitudes to
interfere constructively, and using $\theta_{\eta\eta^\prime} = 20^\circ$. The
experimental upper bounds on the relevant branching ratios are collected in
Appendix~\ref{sec:brarat}. Using these values, we obtain
\beq\label{bounuptetap}
|\xi_{\eta^\prime K_S}| < 0.36\,.
\eeq

So far only upper limits are available for many of the rates that enter in
Eq.~(\ref{boundptetap}).  Hence this bound is probably a significant
overestimate and will improve with further data.  At the present state of the
data, we do not consider it necessary to be concerned about SU(3) breaking
corrections.  Eventually, there may be sufficient data to fix all the
amplitudes $a_f^{u,c}$, including their relative phases. At that point a much
stronger bound can be expected, and allowance for SU(3) breaking corrections
will need to be made.

Using the known CKM dependence, the bound of Eq.~(\ref{bounuptetap}) can be
translated into a bound on the hadronic parameters,
\beq\label{afuafc}
\left|{a^u_{\eta^\prime K^0}\over a^c_{\eta^\prime K^0}}\right| < 18.
\eeq
This bound is much weaker than most theoretical estimates.  Since the
amplitudes involved in Eq.~(\ref{etapks}) carry different strong phases, we do
not expect that they all add up coherently, as assumed in (\ref{boundptetap}).
A more plausible (though less rigorous) estimate would be that the left hand
side of Eq.~(\ref{boundptetap}) is unlikely to be larger than the largest term
on the right hand side. This estimate would give $|\xi_{\eta^\prime K_S}| <
0.14$ instead of $0.36$. Clearly, more data could significantly improve these
bounds.

The same set of SU(3) relationships can be used to carry out a similar analysis
for a number of other modes. The relevant general relationship for
$\eta K_S$ is given in Appendix~\ref{sec:genetapri}. Once experimental data on
the asymmetry in this mode is available, it will be interesting to use
this relationship to obtain a similar constraint on $\xi_{\eta K_S}$.

%%%%%%%%%%%%%%%
%%%%%%%%%%%%%%%
\subsection{Using charged modes and a dynamical assumption}
\label{subsec:etapdyn}

One can obtain a similar bound on the ratio $a_f^u/a_f^c$ for the charged mode,
$f=\eta^\prime K^+$. The experimental situation is such that this bound is
significantly stronger than the one in the neutral mode. The SU(3) relations
for the decompositions of $a_f^u$ and $b_{f'}^u$ in $B^+\to PP$ decays are also
given in Appendix~\ref{sec:suthree}. They lead to the following relations:
\beqa\label{sutchet}
a(\eta_1 K^+)&=&b(\eta_1\pi^+),\no\\
a(\eta_8 K^+)&=&\frac{1}{\sqrt3}\, b(\pi^+\pi^0)-\frac{1}{\sqrt6}\,
b(\overline{K^0}K^+),
\eeqa
and
\beq\label{sutch}
\sqrt6\, b(\eta_8\pi^+)=\sqrt2\, b(\pi^+\pi^0)+2\,
b(\overline{K^0}K^+).
\eeq

This last equation allows us to bound $\xi_{\eta^\prime K^+}$ by many different
combinations of decay modes. The most general relationship that involves only
$B^+$ decay modes is
\beqa\label{etapkp}
a(\eta^\prime K^+)&=&\frac{(3-x)cs}{2}\,
b(\eta\pi^+)+\frac{(x-1)s^2+2c^2}{2}\, b(\eta^\prime\pi^+)\no\\
&&+\frac{(x-3)s}{2\sqrt3}\, b(\pi^+\pi^0) +\frac{xs}{\sqrt6}\,
b(\overline{K^0}K^+),
\eeqa
where, as before, $c$ and $s$ parameterize $\eta-\eta^\prime$ mixing.  The
parameter $x$ is free: it allows us to choose, based on the state of the data,
the optimal (that is, the most constraining) combination of amplitudes. With
the branching ratios collected in Appendix~\ref{sec:brarat}, we find that at
present $x=3$ in Eq.~(\ref{etapkp}) gives the strongest constraint,
\beq\label{bounuptetapc}
|\xi_{\eta^\prime K^+}| < 0.09.
\eeq

While the $a^c_f$ amplitudes for the charged and neutral $\eta^\prime K$ modes
are the same, the $a^u_f$ are not. The non-triplet contributions coming from
the tree $b\to u\bar u d$ terms in the Hamiltonian cause the differences. (This
can be seen in the SU(3) relations in Table~\ref{tab:P8P8} of
Appendix~\ref{sec:suthree}.) If we examine the quark diagrams for these two
channels we find that $a^u_{\eta^\prime K^+}$ has a color-allowed tree diagram
contribution, while $a^u_{\eta^\prime K_S}$ only arises from a color-suppressed
tree diagram or penguins. Our dynamical assumption is that the color-suppressed
$a^u_{\eta^\prime K_S}$ is not bigger than the color-allowed $a^u_{\eta^\prime
K^+}$.  This could only be violated by an accidental cancellation between two
terms that are formally different orders in $1/N_c$.  By making this mild
assumption, we improve the bound on $\xi_{\eta^\prime K_S} $ by more than a
factor of three over that given by the pure SU(3) analysis.

There are experimental tests that could indicate that $a^u_{\eta^\prime K^+}$
is small compared to $a^u_{\eta^\prime K_S}$. First, if direct CP violation in
the neutral mode were established to be large, $|C_{\eta^\prime K_S}|\not\ll1$,
it would place a {\it lower} bound on $|a^u_{\eta^\prime K_S}/a^c_{\eta^\prime
K_S}|$.  Second, if the difference of the neutral and charged $B\to\eta^\prime
K$ rates were sizable, given our strong bound on $|a^u_{\eta^\prime
K^+}/a^c_{\eta^\prime K^+}|$, it would imply a large $|a^u_{\eta^\prime
K_S}/a^c_{\eta^\prime K_S}|$ with a relative strong phase that is not close to
$\pi/2$.  If either of these measurements violate the {\it upper} bound on
$|a^u_{\eta^\prime K^+}/a^c_{\eta^\prime K^+}|$, it would suggest either an
accidental cancellation in the charged mode, or, more interestingly, possible
new physics. This would make it important to improve the direct neutral mode
test for new physics.

Eq.~(\ref{etakwa}) shows that $C_{\eta^\prime K_S}$ is consistent with zero and
the data in Appendix~\ref{sec:brarat} show that the ratio of charged and
neutral rates is not necessarily much different from one. However, this does
not validate our assumption. For example, a small strong phase,
$\delta_{\eta^\prime K}\approx0$, and a large weak phase, $\gamma\approx\pi/2$,
would make direct CP violation small and induce approximately equal rates,
independently of the size of $a_{\eta^\prime K^+}^{u}/a_{\eta^\prime K_S}^u$.

%%%%%%%%%%%%%%%%%%%%%%%%%%%%%%%%%%%%%%%%%%%%%%%%%%%%%%%%%%%%%%%%%%%%%%
\section{\boldmath The CP Asymmetry in $B\to\phi K_S$}
\label{sec:phik}
%%%%%%%%%%%%%%%
%%%%%%%%%%%%%%%
\subsection{SU(3) relations}
\label{subsec:phisut}

A similar analysis can also be applied to $B\to\phi K_S$. Again, the existence
of CKM-suppressed contributions induce a deviation from the leading result, and
our goal is to constrain that effect using SU(3) related modes. Here it is
usually assumed that these corrections are not large, since the $b\to u\bar us$
tree diagram can only contribute via rescattering to the $\phi$ final state
which is pure $s\bar s$.  Thus it was generally argued  that the deviation of
$S_{\phi K_S}$ from $\sin2\beta$ is likely to be no larger than ${\cal
O}(\lambda^2)$. Ref.~\cite{Grossman:1997gr} proposed an SU(3)-based relation
that can potentially bound this deviation, however, it involves an implicit
dynamical assumption. In this subsection we present exact SU(3) relations that
can, in principle, give a model independent bound. In the next subsection, we
explain the dynamical assumption that leads to the bound of
Ref.~\cite{Grossman:1997gr} and update it with current data.

The SU(3) decomposition of $a_f^u$ and $b_{f'}^u$ for final states composed of a
vector and a pseudoscalar meson is given in Appendix~\ref{sec:suthree}. These
results imply the following relations:
\beqa\label{phioe}
a(\phi_1 K^0)&=&-\frac{1}{\sqrt2}\,
b(\phi_1\pi^0)+\sqrt{\frac{3}{2}}\, b(\phi_1\eta_8),\no\\
a(\phi_8 K^0)&=&\frac{1}{4\sqrt2}\, [3b(\rho^0\eta_8)-b(\phi_8\pi^0)]
+\frac{1}{4}\sqrt{\frac{3}{2}}\, [b(\phi_8\eta_8)-b(\rho^0\pi^0)]\no\\
&&+\frac{1}{2}\sqrt{\frac{3}{2}}\,
[b(K^{*0}\overline{K^0})-b(\overline{K^{*0}}K^0)].
\eeqa
The states $\phi_1$ and $\phi_8$ transform as a singlet and an octet of SU(3),
respectively. They are related to the physical $\phi$ and $\omega$ through an
orthogonal rotation:
\beq\label{phiomega}
\phi=\sqrt{\frac{1}{3}}\, \phi_1-\sqrt{\frac{2}{3}}\, \phi_8,\qquad
\omega=\sqrt{\frac{2}{3}}\, \phi_1+\sqrt{\frac{1}{3}}\, \phi_8,
\eeq
which defines the $\phi$ as a pure $s\bar s$ state.
Thus, in terms of physical states, we obtain
\beqa\label{phiks}
a(\phi K^0)&=&
\frac{1}{2}\, [b(\overline{K^{*0}}K^0)-b(K^{*0}\overline{K^0})]
+\frac{1}{2}\sqrt{\frac{3}{2}}\, [cb(\phi\eta)-sb(\phi\eta^\prime)]\no\\
&&+\frac{\sqrt3}{4}\, [cb(\omega\eta)-sb(\omega\eta^\prime)]
-\frac{\sqrt3}{4}\, [cb(\rho^0\eta)-sb(\rho^0\eta^\prime)]\no\\
&&+\frac{1}{4}\, b(\rho^0\pi^0)-\frac{1}{4}b(\omega\pi^0)
-\frac{1}{2\sqrt2}\, b(\phi\pi^0).
\eeqa
This relation could give a bound on $\xi_{\phi K_S}$ in a similar fashion as
Eq.~(\ref{boundptetap}) in Section~\ref{subsec:etapsut}. However, a survey of
the experimental data shows that currently no useful bound can be obtained from
Eq.~(\ref{phiks}).  While it is possible, using SU(3) relations, to replace
some modes that occur on the right-hand side of Eq.~(\ref{phiks}) with a
combination of others, there is no relation that yields a bound on
$\widehat\xi_{\phi K_S}$ below unity at present.

We conclude that, while in the future it will be possible to use relations such
as (\ref{phiks}) to constrain $a^u_{\phi K^0}/a^c_{\phi K^0}$ in a model
independent way, it is impossible to do so with current data.

%%%%%%%%%%%%%%%
%%%%%%%%%%%%%%%
\subsection{Using charged modes and a dynamical assumption}
\label{subsec:phidyn}

For the charged mode $f=\phi K^+$, one can similarly obtain a bound on the
ratio $\xi_{\phi K^+}$  based purely on SU(3). The SU(3) relations for $B^+\to
VP$ decays are given in Appendix~\ref{sec:suthree}. They lead to the following
relations:
\beqa\label{sutchph}
a(\phi_1 K^+)&=&b(\phi_1\pi^+),\no\\
a(\phi_8 K^+)&=&b(\phi_8\pi^+)-\sqrt{\frac{3}{2}}\, b(\overline{K^{*0}}K^+).
\eeqa
In terms of physical states we thus obtain
\beq\label{phikpf}
a(\phi K^+)=b(\phi\pi^+)+b(\overline{K^{*0}}K^+).
\eeq
Using the experimental upper limits on rates collected in
Appendix~\ref{sec:brarat}, we obtain
\beq\label{bounuptphic}
|\xi_{\phi K^+}| < 0.25\,.
\eeq

Once again there is no immediate relationship between the $a^u_{\phi K^+}$ and
$a^u_{\phi K_S}$. They differ by non-triplet Hamiltonian contributions arising
from the tree-type $b \to u\bar u s$ transition (and a small but similar effect
from electroweak penguins). To use the above result as a bound on $\xi_{\phi
K_S}$ requires an additional assumption that $a^u_{\phi K_S}$ is not much
larger than $a^u_{\phi K^+}$. Here we cannot readily justify this assumption,
although we know of no reason why it should not hold. Because the $\phi$ is a
pure $s\bar s$ state, there is no order $N_c^2$ tree contribution to $a^u_{\phi
K^+}$, as there was in the case of $a^u_{\eta K^+}$. The tree $b \to u\bar u s$
contribution must undergo a rescattering in order to contribute. This brings it
to be an order $N_c$ term, at the same level as the penguin contributions.
Traditional analyses of this channel assume that the rescattering contribution
is negligible compared to the $b\to s \bar s s $ penguin terms, in which case
the charged and neutral $a^u_{\phi K}$ would be approximately equal.  Indeed,
only an accidental cancellation could make $|a^u_{\phi K^+}|$ much smaller than
$|a^u_{\phi K_S}|$.

With this assumption, the bound for the charged mode also applies for the
neutral mode. In this case, there is presently no pure SU(3)-based bound, and
so this assumption is necessary to obtain any result. The bound in
Eq.~(\ref{bounuptphic}) was applied to the neutral mode in
Ref.~\cite{Grossman:1997gr}. We have shown here that implicit in this bound is
the assumption that there is no accidental cancellation of the two
contributions to the charged mode.

As in the $\eta^\prime K$ modes, here too there is no evidence either for large
$C_{\phi K_S}$ or for large difference between ${\cal B}(\phi K^0)$ and ${\cal
B}(\phi K^+)$. If such evidence existed, it would indicate that $|a^u_{\phi
K^+}|$ is small compared to $|a^u_{\phi K_S}|$. This could indicate that there
is indeed a cancellation between two independent contributions to $a^u_{\phi
K^+}$, or be a harbinger of new physics effects. Data on the non-strange
neutral modes would then be desirable, as it could distinguish these two
possibilities.

%%%%%%%%%%%%%%%%%%%%%%%%%%%%%%%%%%%%%%%%%%%%%%%%%%%%%%%%%%%%%%%%%%%%%%
\section{\boldmath The CP Asymmetry in $B\to K^+K^-K_S$}
\label{sec:kkk}
The $B\to K^+K^-K_S$ decay is different from the other decay modes
discussed in one, very important, aspect: since the final state is
three-body, it does not have a definite CP. Thus, the CP asymmetry is
diluted with respect to $\sin2\beta$.  To extract the value of
$\sin2\beta$ from this measurement, one has to know the relative
fractions of CP-even and CP-odd final states. The BELLE collaboration
employed a beautiful isospin analysis for this
purpose~\cite{Abe:2002ms}. The accuracy of the isospin analysis
affects the accuracy with which the true $S_{KKK}$ (that is, $S_{KKK}$
for a final $KKK$ state with a definite CP) is
determined. Consequently, the relation between the experimental value,
$S_{KKK}^{\rm exp}$, and $\sin2\beta$ is more complicated.  We will
first analyze the accuracy of determining $S_{KKK}$ from $S_{KKK}^{\rm
exp}$ and then the deviation of $S_{KKK}$ from
$\sin2\beta$.\footnote{In the original version of this section, there
were errors in the isospin and U-spin decompositions of the
relevant amplitudes, which were pointed out in
Ref.~\cite{Gronau:2003ep}. We agree with their results.}

%%%%%%%%%%%%
\subsection{Isospin analysis}
The $B\to K^I K^J K^L$ decays (where $I,J,L = \{+,-,0,S\}$ specify the kaon
states) involve an initial $I=1/2$ state and final $I=1/2$ and $3/2$ states. 
There are five independent isospin amplitudes, $A_1^2$, $A_1^{2'}$, $A_3^2$,
$A_3^{2'}$ and $A_3^4$, where the lower index denotes the isospin
representation of the Hamiltonian and the upper index gives that of the final
state. (We follow the notation of the previous sections, instead of using
isospin labels.) For the isospin-doublet final states, the representation
$2$ denotes where the two $S=-1$ mesons are in an isospin-singlet state, while
$2'$ denotes where they are in an isospin-triplet. Defining
\beq\label{defaijk}
A_{IJL}(p_1,p_2,p_3)\equiv A(B\to K^I(p_1)\overline{K}{}^J(p_2)K^L(p_3)),
\eeq
we obtain the isospin decompositions given in Appendix \ref{sec:isospin}.

Let us start by neglecting the tree contributions, as was done in
the BELLE analysis. This corresponds to $A_3^2=A_3^{2'}=A_3^4=0$. 
Then, the following amplitude relations arise:
\beq\label{eqampl}
A_{00+} = A_{+-0}\,,\qquad
A_{+00} = A_{0-+}\,,\qquad
A_{000} = A_{+-+}\,.
\eeq
When integrating over phase space, the contribution from the 
interference between $A_1^2$ and $A_1^{2'}$ vanishes. 
(Thus, although $A_{IJL}$ is not invariant under $I\leftrightarrow L$,
the rates $\Gamma_{IJL}$ and the branching ratios ${\cal B}_{IJL}$ are.) 
Consequently, the equalities of the following rates are predicted:
\beq\label{eqrates}
\Gamma_{+-0} = \Gamma_{+00}\,,\qquad
\Gamma_{+-+} = \Gamma_{000}\,.
\eeq

Branching ratios of four $B\to KKK$ decays have been 
measured~\cite{Abe:2002ms}:
\beqa\label{expbr}
{\cal B}_{+-+}&=&(3.30\pm0.18\pm0.32)\times10^{-5},\no\\
{\cal B}_{+-0}&=&(2.93\pm0.34\pm0.41)\times10^{-5},\no\\
{\cal B}_{+SS}&=&(1.34\pm0.19\pm0.15)\times10^{-5},\no\\
{\cal B}_{SSS}&=&(0.43^{+0.16}_{-0.14}\pm0.75)\times10^{-5}.
\eeqa
Thus the approximation in Eq.~(\ref{eqampl}) that lead to (\ref{eqrates}) is
not yet tested (in particular, isospin symmetry implies no relation between
${\cal B}_{+-+}$ and ${\cal B}_{+-0}$).  Because $\Gamma_{I00}$ include both
CP-odd and even states for the pair of neutral kaons, the measured rates in
Eq.~(\ref{expbr}) are not sufficient to test the relations in
Eq.~(\ref{eqrates}).  For example, the first relation in Eq.~(\ref{eqrates})
becomes
\beq\label{isotest}
\Gamma_{+-S} = \frac12\, \Gamma_{+SL} + \Gamma_{+SS}\,.
\eeq
 
We now focus on the $K^+K^-K^0$ and $K^0\overline{K^0}K^+$ modes.
One can write down the effective Hamiltonian in terms of the meson
fields. The two $I=0$ terms are of the form $(B^iK_i)(K^jK_j)$ where 
$i$ and $j$ are isospin indices.  We can decompose the Hamiltonian
into components where the $K^jK_j$ pair is either in $l=$ even
or in $l=$ odd angular momentum state:
\beq\label{onlysam}
H_{\rm eff}\propto\, (B^i K_i)\left[x\,(K^j K_j)_{l={\rm even}}
+\sqrt{1-x^2}\,(K^j K_j)_{l={\rm odd}}\right],
\eeq
The equality of the amplitudes in Eq.~(\ref{eqampl}) guarantees that
$x$ is equal for the two decay modes, and allows the extraction of the
CP even/odd fractions in the $B^0\to K_S K^+K^-$ decay from
measurements of the $B^+\to K^+ K_SK_S$ decay as
follows~\cite{Abe:2002ms}.

Consider the $K^+K^-$ sub-system in the $B^0\to K^+ K^-K^0$ decay. Charge
conjugation exchanges $K^+$ and $K^-$, and Parity exchanges them again (in the
center of mass frame). Thus the $K^+K^-$ system has CP$=+1$. Then, the $
K^+K^-K_S$ system has CP$=(-1)^l$, where $l$ is the relative angular momentum
between $K_S$ and $(K^+K^-)$. (It also equals the relative angular momentum
between $K^+$ and $K^-$.) There is then a one-to-one correspondence between
angular momentum and CP. In particular, $x^2$ in Eq.~(\ref{onlysam}) gives
the CP-even fraction in the $B\to K^+K^-K_S$ decay.

Next consider the $\overline{K^0}K^0$ sub-system in the $B^+\to
K^+\overline{K^0}K^0$ decay. Bose symmetry implies that $l=$ even corresponds
to a final $K_SK_S+K_LK_L$ state, while $l=$ odd corresponds to a $K_SK_L$
state. Thus, $x^2=2\Gamma_{+SS}/\Gamma_{+00}$. Since ${\cal B}_{+00}$ has not
been measured, one can use again Eq.~(\ref{eqrates}) to arrive at the following
relation \cite{Abe:2002ms}:
\beq\label{expvalx}
x^2 = 2\,\frac{\Gamma_{+SS}}{\Gamma_{+-0}} = 0.97\pm0.15\pm0.07\,.
\eeq
We learn that the $K^+K^-K_S$ final state is dominantly CP-even. 

From a measured value of the CP asymmetry, $S_{KKK}^{\rm exp}$, we can deduce
the value of the CP asymmetry for the CP-even component, $S_{KKK}$, according
to $S_{KKK}=S_{KKK}^{\rm exp}/(2x^2-1)$. We learn that in the limit that
$I=1$ contributions to the Hamiltonian are neglected, we have
\beq\label{kkkrz}
S_{KKK}=\frac{S_{KKK}^{\rm exp}}{4\Gamma_{+SS}/\Gamma_{+-0} -1}\,.
\eeq

When the higher isospin contributions are taken into account, the
three amplitude equalities of Eq.~(\ref{eqampl}) and the two rate
equalities of Eq.~(\ref{eqrates}) no longer hold. There
remains a single amplitude relation,
\beq\label{exaamre}
A_{000} + A_{+-+} + A_{+00} + A_{00+} + A_{+-0} + A_{0-+} = 0\,.
\eeq
The angular momentum analysis is modified by the three $A_3$ terms.
In particular, both the relation $\Gamma_{+-S}(l={\rm even})/\Gamma_{+-S} =
\Gamma_{+00}(l={\rm even})/\Gamma_{+00}$ and the relation
$\Gamma_{+00}=\Gamma_{+-0}$ are corrected by terms of ${\cal
O}[(A_3^2+\sqrt2A_3^4)/(\sqrt3A_1^2)]$ and of ${\cal
O}[A_3^{2'}/(\sqrt3A_1^{2'})]$. At present there is no experimental 
information on the size of these corrections.

One might worry about isospin violation in the $\phi\to KK$ decays, since
${\cal B}(\phi\to K^+K^-) \approx 49\%$ and ${\cal B}(\phi\to K_SK_L) \approx
34\%$ should be equal in the isospin limit. (This large violation can be
understood as arising chiefly from the phase space difference for the two
channels.)  Since ${\cal B}(B\to \phi K) \times {\cal B}(\phi\to K^+K^-)$ is
between $10-15\%$ of ${\cal B}_{+-0}$, this could give an additional error of
up to $\sim 4\%$ on $x^2$, not a very large effect.

Note that even if the $A_3$ amplitudes were negligibly small and thus
the isospin analysis to find the CP-even fraction in the $KKK$ state
very precise, it would not imply that the extracted CP asymmetry is
equal to $\sin2\beta$ to the same precision. The $b\to u\bar u s$ tree
contribution also has an isosinglet component, which would not affect
the isospin analysis but would shift $S_{KKK}$ from $\sin2\beta$. In
the next subsection we estimate the overall effect of that
contribution.

%%%%%%%%%%%%%%
\subsection{U-spin analysis}
In the previous subsection, we used isospin symmetry to estimate the CP-even
fraction in the $K^+K^-K_S$ final state. Isospin symmetry relates the $B^0\to
K^+K^-K^0$ mode to the $B^+\to K^+K^0\overline{K^0}$ mode.  In this subsection
we use U-spin symmetry to estimate the overall effect of contributions to $B\to
K^+K^-K^+$ that are proportional to $V_{ub}^*V_{us}$. U-spin relates certain
$B^+\to h_i^+h_j^-h_k^+$ modes to each other, where $h_{i,j,k}=K$ or $\pi$.
(Since U-spin is a subgroup of SU(3), this analysis is just a simplified form
of the analysis that we have given for the other channels in this paper.)

Under U-spin, $B^+$ is a singlet, while $M_i=(K^+,\pi^+)$ is a doublet. A
crucial point in our discussion is the U-spin transformation properties of the
Hamiltonian. Both the penguin amplitudes, $b\to(\bar uu+\bar dd+\bar ss)q$, and
the tree contributions $b\to u\bar u q$ (with $q=d,s$) are $\Delta U=1/2$.
Consequently, there are two U-spin amplitudes for the charged $B$ decays
into three charged kaons or pions. The decomposition of the various
decay amplitudes in terms of these two U-spin amplitudes is given in
Table \ref{tab:kkk}. We find the following relation:
\beq\label{equspin}
a(K^+K^-K^+)=b(\pi^+\pi^-\pi^+).
\eeq

The experimental data are \cite{Abe:2002ms,Aubert:2003xz}
\beqa\label{expppp}
{\cal B}_{KKK}\equiv{\cal B}(B^+\to K^+K^-K^+)
&=&(3.1\pm0.2)\times10^{-5},\no\\
{\cal B}_{\pi KK}\equiv{\cal B}(B^+\to \pi^+K^-K^+)
&=&(6.6\pm3.4)\times10^{-6},\no\\
{\cal B}_{K\pi\pi}\equiv{\cal B}(B^+\to K^+\pi^-\pi^+)
&=&(5.7\pm0.4)\times10^{-5},\no\\
{\cal B}_{\pi\pi\pi}\equiv{\cal B}(B^+\to \pi^+\pi^-\pi^+)
&=&(1.1\pm0.4)\times10^{-5}.
\eeqa
To relate the two pairs of rates in a useful way, we make our usual
approximation: we take the $KKK$ rate to be dominated by the
$a_{KKK}^c$ term, and the $\pi\pi\pi$ rate to be dominated by the
$a_{\pi\pi\pi}^u$ term. Then we obtain
\beq\label{treekkk}
|\xi_{KKK}| = \left|\frac{V_{us}}{V_{ud}}\right|
\sqrt{\frac{{\cal B}(B^+\to \pi^+\pi^-\pi^+)} {{\cal
B}(B^+\to K^+K^-K^+)}}\approx0.13,
\eeq
Given the size of U-spin breaking
effects and the crudeness of our approximations, we estimate
that the corrections to $-S_{KKK}=\sin2\beta$ is of the
following size:
\beq\label{estakkk}
\xi_{K^+K^-K_S}=0.13\pm0.06.
\eeq

Additional constraint on $|\xi_{KKK}|$ can be derived from ${\cal
B}_{\pi KK}$. Our amplitude relations imply that, in the U-spin limit,
${\cal B}_{\pi KK}\geq{\cal B}_{\pi\pi\pi}/2$. Consequently,
\beq\label{treekkp}
|\xi_{KKK}| \leq \left|\frac{V_{us}}{V_{ud}}\right|
\sqrt{\frac{2{\cal B}(B^+\to \pi^+K^-K^+)} {{\cal
B}(B^+\to K^+K^-K^+)}}\approx0.14.
\eeq

The above analysis does not distinguish between quasi-two-body and true
three-body contributions to the eventual three-body rate. Indeed it includes
all three-body final states, whether or not reached by a resonant contribution.
We used here the SU(3) relationship only for the total rates, integrated over
the entire Dalitz plots. Comparisons of more restricted regions of the Dalitz
plots would be much more subject to SU(3) breaking corrections.

%%%%%%%%%%%%%%%%%%%%%%%%%%%%%%%%%%%%%%%%%%%%%%%%%%%%%%%%%%%%%%%%%%%%%%
\section{Conclusions}
\label{sec:con}

Within the Standard Model, the CP asymmetries $S_f$ in neutral $B$
decays to the final CP eigenstates $\phi K_S$, $\eta^\prime K_S$ and
$(K^+K^-K_S)_{{\rm CP}=-1}$ are equal to the CKM parameter
$\sin2\beta$ measured in $B\to \psi K_S$, to a good
approximation. Furthermore, the direct CP asymmetries $C_f$ in these
modes are expected to be small.  The goodness of this approximation is
different between the various modes and its estimate suffers, in
general, from hadronic uncertainties. We used SU(3) relations and
experimental data (and, in some cases, a mild dynamical assumption) to
estimate or to derive upper bounds on the deviation of the $S_f$
from $\sin2\beta$ and on the size of $C_f$. We obtained
\beqa\label{sumxif}
|\xi_{\eta^\prime K_S}| &<& \cases{0.36 & SU(3),\cr
0.09 & \mbox{SU(3) + leading $N_c$ assumption},\cr}\no\\[6pt]
|\xi_{\phi K_S}| &<&\ \ \, 0.25\ \ \
\mbox{SU(3) + non-cancellation assumption},\no\\
|\xi_{K^+K^-K_S}| &\sim&\ \ \, 0.13\ \ \ \,\mbox{U-spin},
\eeqa
where $\xi_f$ is defined in Eq.~(\ref{defxi}). The approximations and
assumptions that lead to these results are spelled out in the
corresponding sections. While our bounds for the first two modes are
considerably weaker than estimates based on explicit calculations of
the hadronic amplitudes, they have the advantage that they are model
independent. Although SU(3) breaking effects could be significant, our
bounds for the two-body modes are probably still conservative because
they arise from a sum over several complex amplitudes that we assumed
to interfere constructively. Furthermore only experimental upper
bounds are available for many of the rates that enter these bounds. As
data improves these bounds could become significantly stronger.
Certainly, if deviations from $\sin2\beta$ are established that are
larger than the SU(3) bounds, the case for new physics would be
convincing. Since our bounds apply more generally to minimal flavor
violation models, the new physics would have to be beyond this framework.
Even where our results require additional assumptions, the situation
here is better than the usual, in that we are making assumptions about
non-leading corrections to the $a^u_f$ amplitudes, rather than about
the full $a^u_f$ terms.

%%%%%%%%%%%%
\acknowledgments
We thank Adam Falk and Adam Schwimmer for helpful discussions, and the CERN and
SLAC theory groups for hospitality while parts of this work were completed.
The work of Y.G., Z.L. and H.Q. was supported in part by the United
States--Israel Binational Science Foundation through Grant No.~2000133.
The research of Y.G. and Y.N. is supported by a Grant from the G.I.F.,
the German--Israeli Foundation for Scientific Research and
Development, and by the Einstein Minerva Center for Theoretical Physics.
Y.G. was supported in part by the Israel Science Foundation under
Grant No.~237/01.
Z.L.\ was supported in part by the Director, Office of Science, Office of
High Energy and Nuclear Physics, Division of High Energy Physics, of the  U.S.\
Department of Energy under Contract DE-AC03-76SF00098 and by a DOE Outstanding
Junior Investigator award.
Y.N.\ is supported by the Israel Science Foundation founded by the Israel
Academy of Sciences and Humanities and by EEC RTN contract HPRN-CT-00292-2002.
The work of H.Q.\ was supported by the Department of Energy, contract
DE-AC03-76SF00515.

%%%%%%%%%%%%%%%%%%%%%%%%%%%%%%%%%%%%
\appendix
\section{\boldmath SU(3) decomposition for
  $\langle f^{(\prime)}|(\bar bu)(\bar uq)|B\rangle$ amplitudes}
\label{sec:suthree}

In this appendix we give the SU(3) decomposition of matrix elements that are
relevant to our analysis.  The operator that creates a $B$ meson containing a
$\bar b$ quark transforms as a $\overline{3}$ of SU(3).  The $\Delta B = +1$
Hamiltonian, which has the flavor structure $(\bar b\, q_i)(\bar q_j q_k)$,
transforms as $3 \times 3 \times \overline{3} = 15 + \overline{6} + 3 + 3$.
Our calculations follow closely that in Ref.~\cite{Falk:2001hx}, the only
difference being the decomposition of the Hamiltonian that can be read off from
Ref.~\cite{Savage:ub}.

\begin{table}[hb]
\caption{SU(3) decomposition of $a(f)$ and $b(f')$ for
$f^{(\prime)} = \eta_1 P_8$.}
\label{tab:1P8}
\begin{tabular}{c||ccc} \hline\hline
$f^{(\prime)}$  &  $S_{15}^8$  &  $S_{\overline{6}}^8$  &  $S_3^8$ \\[2pt] \hline\hline
$\eta_1 K^0$  &  $-1$  &  $-1$  &  $1$ \\ \hline
$\eta_1 K^+$  &  $3$  &  $1$   &  $1$  \\ \hline\hline
$\eta_1\pi^0$  &  $5/\sqrt2$  &  $-1/\sqrt2$  &  $-1/\sqrt2$ \\
$\eta_1\eta_8$  &  $\sqrt{3/2}$  &  $-\sqrt{3/2}$  &  $1/\sqrt{6}$ \\ \hline
$\eta_1\pi^+$ &  $3$  &  $1$  &  $1$
\\ \hline\hline
\end{tabular}
\end{table}

For final states composed of an SU(3) singlet and an octet, there are three
reduced matrix elements. The reason is that there is a unique way of making a
singlet from an octet plus any one of the three representations of the
Hamiltonian and the $B$ operator.  In Table~\ref{tab:1P8} we give the
decomposition of $a(f) \equiv a_f^{u,c}$ and $b(f^\prime) \equiv b_{f'}^{u,c}$
for $f^{(\prime)} = \eta_1 P_8$, where $\eta_1$ is the SU(3)-singlet
pseudoscalar, and $P_8$ is the SU(3)-octet pseudoscalar.  The matrix elements
$S_\alpha^\beta$ that occur in the decomposition of $a_f^u$ and $b_{f'}^u$ are
independent of those for $a_f^c$ and $b_{f'}^c$.  In our notation, the lower
index of the reduced matrix elements denotes the SU(3) representation of the
Hamiltonian and the upper index is that of the final state.  (If electroweak
penguin contributions were neglected, the decomposition of $a_f^c$ and
$b_{f'}^c$ is given by the last column, corresponding to $H$ in a triplet.) The
decomposition for $f^{(\prime)} = \phi_1 P_8$, where $\phi_1$ is the
SU(3)-singlet vector-meson, is the same as that for $f^{(\prime)} = \eta_1
P_8$, with different values of the reduced matrix elements, $S_\alpha^\beta$,
and the replacement $\eta_1\to\phi_1$.

\begin{table}[t]
\caption{SU(3) decomposition of $a(f)$ and $b(f')$ for
$f^{(\prime)} = P_8 P_8$.}
\label{tab:P8P8}
\begin{tabular}{c||ccccc} \hline\hline
$f^{(\prime)}$  &  $A_{15}^{27}$  &  $A_{15}^8$  &  $A_{\overline{6}}^8$
&  $A_{3}^8$ & $A_3^1$
\\[2pt] \hline\hline
$\eta_8 K^0$  &  $4 \sqrt{6}/5$  &  $1/\sqrt{6}$  &  $-1/\sqrt{6}$
&  $-1/\sqrt{6}$  &  $0$ \\
$K^0 \pi^0$  &  $12 \sqrt{2}/5$  &  $1/\sqrt{2}$  &  $-1/\sqrt{2}$
&  $-1/\sqrt{2}$  &  $0$ \\
$K^+ \pi^-$  &  $16/5$  &  $-1$  &  $1$  &  $1$  &  $0$ \\ \hline
$\eta_8 K^+$  &  $8 \sqrt{6}/5$  &  $-\sqrt{3/2}$  &  $1/\sqrt{6}$
&  $-1/\sqrt{6}$  &  $0$ \\
$K^+ \pi^0$  &  $16 \sqrt{2}/5$  &  $3/\sqrt{2}$  &  $-1/\sqrt{2}$
&  $1/\sqrt{2}$  &  $0$ \\
$K^0 \pi^+$  &  $-8/5$  &  $3$  &  $-1$  &  $1$ & $0$ \\ \hline\hline
$\eta_8 \pi^0$  &  $0$  &  $5/\sqrt{3}$  &  $1/\sqrt{3}$  &  $-1/\sqrt{3}$
&  $0$ \\
$\pi^0 \pi^0$  &  $-13\sqrt2/5$  &  $1/\sqrt2$  &  $1/\sqrt2$
&  $1/(3\sqrt2)$  &  $\sqrt2$ \\
$\eta_8 \eta_8$  &  $3\sqrt2/5$  &  $-1/\sqrt2$  &  $-1/\sqrt2$
&  $-1/(3\sqrt2)$  &  $\sqrt2$ \\
$\pi^- \pi^+$  &  $14/5$  &  $1$  &  $1$  &  $1/3$  &  $2$ \\
$K^- K^+$  &  $-2/5$  &  $2$  &  $0$  &  $-2/3$  &  $2$ \\
$K^0 \overline{K^0}$  &  $-2/5$  &  $-3$  &  $-1$  &  $1/3$  &  $2$ \\ \hline
$\eta_8 \pi^+$  &  $4 \sqrt{6}/5$  &  $\sqrt{6}$  &  $-\sqrt{2/3}$  
&  $\sqrt{2/3}$  &  $0$ \\
$\pi^+ \pi^0$  &  $4 \sqrt{2}$  &  $0$  &  $0$  &  $0$  &  $0$ \\
$K^+ \overline{K^0}$  &  $-8/5$  &  $3$  &  $-1$  &  $1$  &  $0$
\\ \hline\hline
\end{tabular}
\end{table}

Final states containing two SU(3) octets can be decomposed as $8 \times 8 = 27
+ 10 + \overline{10} + 8_S + 8_A + 1$.  The final state composed of $P_8 P_8$
is symmetric, and so it transforms as an element of the symmetric part, $(8
\times 8)_S = 27 + 8 + 1$. In Table~\ref{tab:P8P8} we give the decomposition of
$a(f)$ and $b(f)$ for $f^{(\prime)} = P_8 P_8$; it contains five reduced matrix
elements.  When the final mesons are different, such as $f^{(\prime)} = P_8
V_8$, where $V_8$ is the SU(3)-octet vector-meson, all six representations
appear.  In Table~\ref{tab:V8P8} we give the decomposition of $a(f)$ and
$b(f')$ for $f^{(\prime)} = P_8 V_8$, which contains ten reduced matrix
elements.  Again, the matrix elements $A_\alpha^\beta$ and $B_\alpha^\beta$
that occur in the decomposition of $a_f^u$ and $b_{f'}^u$ are independent of
those for $a_f^c$ and $b_{f'}^c$. (If electroweak penguins were neglected, the
decomposition of $a_f^c$ and $b_{f'}^c$ is given by the columns corresponding
to $H$ in a triplet, $A_3^8$, $A_3^1$, $B_3^{8_S}$, $B_3^{8_A}$, and $B_3^1$.)

\begingroup\squeezetable
\begin{table}[t]
\caption{SU(3) decomposition of $a(f)$ and $b(f')$ for
$f^{(\prime)} = V_8 P_8$.}
\label{tab:V8P8}
\begin{tabular}{c||cccccccccc} \hline\hline
$f^{(\prime)}$  &  $B_{15}^{27}$  &  $B_{15}^{8_S}$  &  $B_{\overline{6}}^{8_S}$  &  $B_3^{8_S}$ &  $B_{3}^1$
  &  $B_{15}^{10}$  &  $B_{\overline{6}}^{\overline{10}}$  &  $B_{15}^{8_A}$   &  $B_{\overline{6}}^{8_A}$
  &  $B_3^{8_A}$
\\[2pt] \hline\hline
$\phi_8 K^0$  &  $2 \sqrt{6}/5$  &  $1/(2 \sqrt{6})$  &  $-1/(2 \sqrt{6})$  &  $-1/(2 \sqrt{6})$  &  $0$&
  $-4 \sqrt{2/3}$  &  $0$  &  $\sqrt{3/2}/2$  &  $-\sqrt{3/2}/2$  &  $-\sqrt{3/2}/2$ \\
$K^{*0} \eta_8$  &  $2 \sqrt{6}/5$  &  $1/(2 \sqrt{6})$  &  $-1/(2
  \sqrt{6})$  &  $-1/(2 \sqrt{6})$  & $0$  &  $ 4 \sqrt{2/3}$  &  $0$  &  $-\sqrt{3/2}/2$  &  $\sqrt{3/2}/2$  &  $\sqrt{3/2}/2$ \\
$K^{*0} \pi^0$  &  $6 \sqrt{2}/5$  &  $1/(2 \sqrt{2})$  &  $-1/(2 \sqrt{2})$  &  $-1/(2 \sqrt{2})$  &  $0$  &
  $ 4 \sqrt{2}/3$  &  $4 \sqrt{2}/3$  &  $1/(2 \sqrt{2})$  &  $-1/(2 \sqrt{2})$  &  $-1/(2 \sqrt{2})$ \\
$\rho^0 K^0$  &  $6 \sqrt{2}/5$  &  $1/(2 \sqrt{2})$  &  $-1/(2 \sqrt{2})$  &  $-1/(2 \sqrt{2})$  &  $0$  &
  $-4 \sqrt{2}/3$  &  $-4 \sqrt{2}/3$  &  $-1/(2 \sqrt{2})$  &  $1/(2 \sqrt{2})$  &  $1/(2 \sqrt{2})$ \\
$K^{*+} \pi^-$  &  $8/5$  &  $-1/2$  &  $1/2$  &  $1/2$  &  $0$  & $-8/3$  &  $4/3$  &  $-1/2$  &  $1/2$  &  $1/2$ \\
$\rho^- K^+$  &  $8/5$  &  $-1/2$  &  $1/2$  &  $1/2$  &  $0$  &
  $8/3$  &  $-4/3$  &  $1/2$  &  $-1/2$  &  $-1/2$ \\
\hline
$\phi_8 K^+$  &  $4 \sqrt{6}/5$  &  $-\sqrt{3/2}/2$  &  $1/(2 \sqrt{6})$  &  $-1/(2 \sqrt{6})$  &  $0$&
  $0$  &  $0$  &  $-3 \sqrt{3/2}/2$  &  $\sqrt{3/2}/2$  &  $-\sqrt{3/2}/2$ \\
$K^{*+} \eta_8$  &  $4 \sqrt{6}/5$  &  $-\sqrt{3/2}/2$  &  $1/(2 \sqrt{6})$  &  $-1/(2 \sqrt{6})$  &  $0$  &  $0$  &  $0$  &
  $3 \sqrt{3/2}/2$  &  $-\sqrt{3/2}/2$  &  $\sqrt{3/2}/2$ \\
$K^{*+} \pi^0$  &  $8 \sqrt{2}/5$  &  $3/(2 \sqrt{2})$  &  $-1/(2 \sqrt{2})$  &  $1/(2 \sqrt{2})$  &  $0$  &
  $0$  &  $4 \sqrt{2}/3$  &  $3/(2 \sqrt{2})$  &  $-1/(2 \sqrt{2})$  &  $1/(2 \sqrt{2})$ \\
$\rho^0 K^+$  &  $8 \sqrt{2}/5$  &  $3/(2 \sqrt{2})$  &  $-1/(2 \sqrt{2})$  &  $1/(2 \sqrt{2})$  &  $0$  &
  $0$  &  $-4 \sqrt{2}/3$  &  $-3/(2 \sqrt{2})$  &  $1/(2 \sqrt{2})$  &  $-1/(2 \sqrt{2})$ \\
$K^{*0} \pi^+$  &  $-4/5$  &  $3/2$  &  $-1/2$  &  $1/2$  &  $0$&
  $0$  &  $-4/3$  &  $3/2$  &  $-1/2$  &  $1/2$ \\
$\rho^+ K^0$  &  $-4/5$  &  $3/2$  &  $-1/2$  &  $1/2$  &  $0$  &  $0$  &  $4/3$  &  $-3/2$  &  $1/2$  &  $-1/2$ \\
\hline\hline
$\phi_8 \pi^0$  &  $0$  &  $5/(2 \sqrt{3})$  &  $1/(2 \sqrt{3})$  &  $-1/(2 \sqrt{3})$  &  $0$  &  $4/\sqrt{3}$  &  $2/\sqrt{3}$  &  $0$  &  $0$  &  $0$ \\
$\rho^0 \eta_8$  &  $0$  &  $5/(2 \sqrt{3})$  &  $1/(2 \sqrt{3})$  &  $-1/(2 \sqrt{3})$  &  $0$  &  $-4/\sqrt{3}$  &  $-2/\sqrt{3}$  &  $0$  &  $0$  &  $0$ \\
$\rho^0 \pi^0$  &  $-13/5$  &  $1/2$  &  $1/2$  &  $1/6$  &  $1$  &  $0$  &  $0$  &  $0$  &  $0$  &  $0$ \\
$\phi_8 \eta_8$  &  $3/5$  &  $-1/2$  &  $-1/2$  &  $-1/6$  &  $1$  &  $0$  &  $0$  &  $0$  &  $0$  &  $0$ \\
$\rho^- \pi^+$  &  $7/5$  &  $1/2$  &  $1/2$  &  $1/6$  &  $1$  &  $4/3$  &  $-2/3$  &  $5/2$  &  $1/2$  &  $-1/2$ \\
$\rho^+ \pi^-$  &  $7/5$  &  $1/2$  &  $1/2$  &  $1/6$  &  $1$  &  $-4/3$  &  $2/3$  &  $-5/2$  &  $-1/2$  &  $1/2$ \\
$K^{*-} K^+$  &  $-1/5$  &  $1$  &  $0$  &  $-1/3$  &  $1$  &  $-4/3$  &  $2/3$  &  $2$  &  $1$  &  $0$ \\
$K^{*+} K^-$  &  $-1/5$  &  $1$  &  $0$  &  $-1/3$  &  $1$  &  $4/3$  &  $-2/3$  &  $-2$  &  $-1$  &  $0$ \\
$K^{*0} \overline{K^0}$  &  $-1/5$  &  $-3/2$  &  $-1/2$  &  $1/6$  &  $1$  &  $-4/3$  &  $2/3$  &  $1/2$  &  $-1/2$  &  $-1/2$ \\
$\overline{K^{*0}} K^0$  &  $-1/5$  &  $-3/2$  &  $-1/2$  &  $1/6$  &  $1$  &  $4/3$  &  $-2/3$  &  $-1/2$  &  $1/2$  &  $1/2$ \\
\hline
$\phi_8 \pi^+$  &  $2 \sqrt{6}/5$  &  $\sqrt{3/2}$  &  $-1/\sqrt{6}$  &  $1/\sqrt{6}$  &  $0$  &  $0$  &  $-2 \sqrt{2/3}$  &  $0$  &  $0$  &  $0$ \\
$\rho^+ \eta_8$  &  $2 \sqrt{6}/5$  &  $\sqrt{3/2}$  &  $-1/\sqrt{6}$  &  $1/\sqrt{6}$  &  $0$  &  $0$  &  $2 \sqrt{2/3}$  &  $0$  &  $0$  &  $0$ \\
$\rho^+ \pi^0$  &  $2 \sqrt{2}$  &  $0$  &  $0$  &  $0$  &  $0$  &  $0$  &  $2 \sqrt{2}/3$  &  $3/\sqrt{2}$  &  $-1/\sqrt{2}$  &  $1/\sqrt{2}$ \\
$\rho^0 \pi^+$  &  $2 \sqrt{2}$  &  $0$  &  $0$  &  $0$  &  $0$  &  $0$  &  $-2 \sqrt{2}/3$  &  $-3/\sqrt{2}$  &  $1/\sqrt{2}$  &  $-1/\sqrt{2}$ \\
$\overline{K^{*0}} K^+$  &  $-4/5$  &  $3/2$  &  $-1/2$  &  $1/2$  &  $0$  &  $0$  &  $-4/3$  &  $3/2$  &  $-1/2$  &  $1/2$ \\
$K^{*+} \overline{K^0}$  &  $-4/5$  &  $3/2$  &  $-1/2$  &  $1/2$  &  $0$  &  $0$  &  $4/3$  &  $-3/2$  &  $1/2$  &  $-1/2$
\\ \hline\hline
\end{tabular}
\end{table}
\endgroup

Related tables have been presented in Ref.~\cite{Savage:ub}, however, the
$(\bar b u)(\bar us)$ contributions to strangeness changing decays were
neglected. Furthermore, in that paper, when there are several independent
amplitudes with the Hamiltonian in a given SU(3) representation, these
contributions are not decomposed according to the SU(3) representation of the
final state. Ref.~\cite{Grinstein:1996us} gives the SU(3) decomposition for
$f^{(\prime)} = P_8P_8$ in a somewhat different notation from ours.  They do
not discuss the applications we investigate here. In
Ref.~\cite{Deshpande:2000jp} a nonet [U(3)] symmetry is assumed (which can be
justified in the large $N_c$ limit).  It relates matrix elements involving
$\phi_1$ and $\phi_8$ (and, similarly, $\eta_1$ and $\eta_8$) that are
independent of one another based only on SU(3).  Thus, in
Ref.~\cite{Deshpande:2000jp}, eleven amplitudes describe $B\to VP$ decays (with
$V$ in a singlet or octet and $P$ in an octet), while we need thirteen. The use
of the fewer number of matrix elements amounts to a dynamical assumption beyond
SU(3).

%%%%%%%%%%%%%%%%%%%%%%%%%%%%%%%%%%%%
\section{\boldmath SU(3) relations for $B^0\to\eta^\prime K^0$ and $B^0\to\eta K^0$}
\label{sec:genetapri}

The most general SU(3) relation between the $a(\eta^\prime K^0)$ and
$b(f^\prime)$'s of charged and neutral $B$ decays can be written as follows:
\beqa\label{genetapk}
a(\eta^\prime K^0)&=&
\left[\frac{s^2-2c^2}{2\sqrt2}-\frac{\sqrt3s^2(x_1-x_2)}{2}\right]
b(\eta^\prime\pi^0)
-\left[\frac{3sc}{2\sqrt2}-\frac{\sqrt3sc(x_1-x_2)}{2}\right]
b(\eta\pi^0)\no\\
&+&\left[\frac{\sqrt3s}{4}+\frac{s(x_1+x_2+4x_3)}{2\sqrt2}\right]
b(\pi^0\pi^0)
-\left[\frac{\sqrt{3}s(s^2+4c^2)}{4}-\frac{3s^3(x_1+x_2)}{2\sqrt2}\right]
b(\eta^\prime\eta^\prime)\no\\
&+&\left[\frac{3\sqrt{3}sc^2}{4}
+\frac{3 sc^2(x_1+x_2)}{2 \sqrt2}\right]
b(\eta\eta)
+\left[\frac{\sqrt{6}c(2c^2-s^2)}{4}-\frac{3cs^2(x_1+x_2)}{2}\right]
b(\eta\eta^\prime)\no\\
&-&sx_3\, b(\pi^+\pi^-)-sx_1\, b(K^+K^-)-sx_2\, b(K^0\overline{K^0})
-sx_4\, b(\overline{K^0}K^+)\no\\
&+&\sqrt{\frac{3}{2}}\, scx_4\, b(\eta\pi^+)
-\sqrt{\frac{3}{2}}\, s^2x_4\, b(\eta^\prime\pi^+)
+\left(\sqrt2sx_3-\frac{sx_4}{\sqrt2}\right)b(\pi^0\pi^+).
\eeqa
Here, $x_1$, $x_2$, $x_3$ and $x_4$ are free parameters that allow us to choose
between various combinations of amplitudes on the right hand side of this
relation. In particular, given a set of experimental measurements of (or bounds
on) the corresponding branching ratios, we can vary the $x_i$ parameters so
that we get the strongest constraint. With current data, the optimal choice is
$x_1=x_2=x_3=x_4=0$, which yields Eq.~(\ref{etapks}).

The analogous relation that will allow to bound the deviation of the CP
asymmetry in $B\to \eta K_S$ decay (once it is measured) from $\sin2\beta$ is:
\beqa\label{etakzci}
a(\eta K^0)&=&-\frac{sc}{2} \left(\sqrt3x_1+\frac{3}{\sqrt2}\right)
b(\eta^\prime\pi^0)-\left[\frac{s^2}{\sqrt2}-\frac{c^2}{2}\left(\sqrt3x_1+\frac{1}{\sqrt2}\right)
\right] b(\eta\pi^0)\no\\
&+&\sqrt{\frac32}\,scx_2\, b(\eta^\prime\pi^+)
-\sqrt{\frac32}\,c^2x_2\, b(\eta\pi^+)
+\frac{s}2\left(c^2\left(\sqrt{\frac32}+3x_3\right)-
s^2\sqrt6\right) b(\eta^\prime\eta)\no\\
&-&\frac{3s^2c}4\left(\sqrt3+\sqrt2 x_3\right) b(\eta^\prime\eta^\prime)
+c\left[\frac{c^2}{4}\left(\sqrt3-3\sqrt2 x_3\right)+s^2\sqrt{3}\right]b(\eta\eta)
\no\\
&-&\frac{c}{4}\left(\sqrt3+\sqrt2(4x_4+x_3)\right)b(\pi^0\pi^0)+
cx_4\,b(\pi^+\pi^-)
  +\frac{c}{\sqrt2}\left(x_2-2x_4\right) b(\pi^0\pi^+)\no\\
&+&\frac{c}{2}\left(x_3-x_1\right) b(K^+K^-)
+\frac{c}{2}\left(x_3+x_1\right)b(K^0\overline{K^0})
+cx_2\,b(\overline{K^0}K^+)\,.
\eeqa

We have also derived the most general SU(3) relation between the $a(\phi_8
K^0)$ and the (sixteen) $b_{f'}^u$'s of charged and neutral $B$ decays. The
relation is quite complicated and it does not seem likely that it will become
useful in the near future, so we do not present it explicitly here.

%%%%%%%%%%%%%%%%%%%%%%%%%%%%%%%%%%%%
\section{Relevant Branching Ratios}
\label{sec:brarat}

The values below are collected from Ref.~\cite{Hagiwara:fs}.
\beqa\label{pdgetak}
{\cal B}(\eta^\prime K^0)&=&(5.8^{+1.4}_{-1.3})\times10^{-5},\no\\
{\cal B}(\eta^\prime K^+)&=&(7.5\pm0.7)\times10^{-5},\no\\
{\cal B}(\pi^+\pi^-)&=&(4.4\pm0.9)\times10^{-6},\no\\
{\cal B}(K^+K^-)&<&1.9\times10^{-6},\no\\
{\cal B}(\overline{K^0}K^+)&<&2.4\times10^{-6},\no\\
{\cal B}(\eta\pi^0)&<&2.9\times10^{-6},\no\\
{\cal B}(\pi^0\pi^0)&<&5.7\times10^{-6},\no\\
{\cal B}(\eta\pi^+)&<&5.7\times10^{-6},\no\\
{\cal B}(\eta^\prime\pi^0)&<&5.7\times10^{-6},\no\\
{\cal B}(\eta^\prime\pi^+)&<&7.0\times10^{-6},\no\\
{\cal B}(\pi^0\pi^+)&<&9.6\times10^{-6},\no\\
{\cal B}(K^0\overline{K^0})&<&1.7\times10^{-5},\no\\
{\cal B}(\eta\eta)&<&1.8\times10^{-5},\no\\
{\cal B}(\eta\eta^\prime)&<&2.7\times10^{-5},\no\\
{\cal B}(\eta^\prime\eta^\prime)&<&4.7\times10^{-5}.
\eeqa
\beqa\label{pdgphik}
{\cal B}(\phi K^0)&=&(8.1^{+3.2}_{-2.6})\times10^{-6},\no\\
{\cal B}(\phi K^+)&=&(7.9^{+2.0}_{-1.8})\times10^{-6},\no\\
{\cal B}(\eta\omega)&<&1.2\times10^{-5},\no\\
{\cal B}(\eta^\prime\omega)&<&6.0\times10^{-5},\no\\
{\cal B}(\eta\phi)&<&0.9\times10^{-5},\no\\
{\cal B}(\eta^\prime\phi)&<&3.1\times10^{-5},\no\\
{\cal B}(\eta\rho^0)&<&1.0\times10^{-5},\no\\
{\cal B}(\eta^\prime\rho^0)&<&1.2\times10^{-5},\no\\
{\cal B}(\rho^0\pi^0)&<&5.5\times10^{-6},\no\\
{\cal B}(\omega\pi^0)&<&3.0\times10^{-6},\no\\
{\cal B}(K^+\overline{K^{*0}})&<&5.3\times10^{-6},\no\\
{\cal B}(\phi\pi^+)&<&1.4\times10^{-6}.
\eeqa

%%%%%%%%%%%%%%%%%%%%%%%%%%%%%%%%%%%%
\section{\boldmath SU(2) decomposition for
  $\langle KKK|(\bar bu)(\bar us)|B\rangle$ amplitudes}
\label{sec:isospin}

In this appendix we give the isospin decomposition of matrix elements that are
relevant to our analysis.  The operator that creates a $B$ meson containing a
$\bar b$ quark transforms as a ${2}$ of isospin-SU(2).  The $\Delta B = +1$
Hamiltonian, which has the flavor structure $(\bar b\, s)(\bar q_i q_i)$,
transforms as either $2 \times 2 = 1 + 3$ for $q_i=u,d$ or $1$ for $q_i=s$.
For final isospin-doublet states, the non-primed (primed) isospin amplitudes
correspond to the $K^IK^L$ sub-system being in an isospin-singlet
(-triplet) state.
A similar form of decomposition holds for the U-spin amplitudes in
$B^+\to h_i^+(p_1) h_j^-(p_2) h_l^+(p_3)$ decays (where $h_{i,l}^+=K^+,\pi^+$
and $h_j^-=\pi^-,K^-$). They are given in Table \ref{tab:uspin}.

\begin{table}[t]
\caption{Left: Isospin decomposition of $A_{IJL}=A(B\to
K^I(p_1)\overline{K}{}^J(p_2)K^L(p_3))$. Right: 
U-spin decomposition of $A_{h_ih_jh_l}=A(B\to
h_i^+(p_1)h_j^-(p_2)h_l^+(p_3))$.} 
\label{tab:kkk}
\begin{tabular}{c||ccccc} \hline\hline
$A_{IJL}$  &  $A_1^2$  &  $A_1^{2'}$  &  $A_3^2$ &
$A_3^{2'}$ & $A_3^4$ \\[2pt] \hline\hline
$A_{+00}$  &  $\frac{1}{2\sqrt3}$  &  $-\frac12$  &  $\frac16$ &
$-\frac{1}{2\sqrt3}$ & $\frac{1}{3\sqrt2}$ \\
$A_{00+}$  &  $\frac{1}{2\sqrt3}$  &  $\frac12$  &  $\frac16$ &
$\frac{1}{2\sqrt3}$ & $\frac{1}{3\sqrt2}$ \\
$A_{+-0}$  &  $\frac{1}{2\sqrt3}$  &  $\frac12$  &  $-\frac16$ &
$-\frac{1}{2\sqrt3}$ & $-\frac{1}{3\sqrt2}$ \\
$A_{0-+}$  &  $\frac{1}{2\sqrt3}$  &  $-\frac12$  &  $-\frac16$ &
$\frac{1}{2\sqrt3}$ & $-\frac{1}{3\sqrt2}$ \\
$A_{000}$  &  $-\frac{1}{\sqrt3}$  &  $0$  &  $\frac13$ &
$0$ & $-\frac{1}{3\sqrt2}$ \\
$A_{+-+}$  &  $-\frac{1}{\sqrt3}$  &  $0$  &  $-\frac13$ &
$0$ & $\frac{1}{3\sqrt2}$  
\\[2pt] \hline\hline
\end{tabular}
\hspace*{2cm}
\label{tab:uspin}
\begin{tabular}{c||cc} \hline\hline
$A_{h_ih_lh_j}$  &  $X_1^2$  &  $X_1^{2'}$ \\[2pt] \hline\hline
$A_{K\pi\pi}$  &  $\frac{1}{2\sqrt3}$  &  $-\frac12$ \\
$A_{\pi\pi K}$  &  $\frac{1}{2\sqrt3}$  &  $\frac12$ \\
$A_{KK\pi}$  &  $\frac{1}{2\sqrt3}$  &  $\frac12$  \\
$A_{\pi KK}$  &  $\frac{1}{2\sqrt3}$  &  $-\frac12$ \\
$A_{\pi\pi\pi}$  &  $-\frac{1}{\sqrt3}$  &  $0$ \\
$A_{KKK}$  &  $-\frac{1}{\sqrt3}$  &  $0$  
\\[2pt] \hline\hline
\end{tabular}
\end{table}

\newpage

%%%%%%%%%%%%%%%%%%%%%%%%%%%


\begin{thebibliography}{13}

%\cite{Grossman:1996ke}
\bibitem{Grossman:1996ke}
Y.~Grossman and M.~P.~Worah,
%``CP asymmetries in B decays with new physics in decay amplitudes,''
Phys.\ Lett.\ B {\bf 395}, 241 (1997)
[hep-ph/9612269].
%%CITATION = HEP-PH 9612269;%%

%\cite{Nir:2002gu}
\bibitem{Nir:2002gu}
Y.~Nir,
%``CP violation: The CKM matrix and new physics,''
hep-ph/0208080.
%%CITATION = HEP-PH 0208080;%%

%\cite{Aubert:2002ic}
\bibitem{Aubert:2002ic}
B.~Aubert {\it et al.}, BABAR Collaboration,
%``Measurement of the CP-violating asymmetry amplitude sin 2beta,''
Phys.\ Rev.\ Lett.\  {\bf 89}, 201802 (2002)
[hep-ex/0207042].
%%CITATION = HEP-EX 0207042;%%

%\cite{Abe:2002px}
\bibitem{Abe:2002px}
K.~Abe {\it et al.}, BELLE Collaboration,
%``An improved measurement of mixing-induced CP violation in the neutral B  meson system,''
Phys.\ Rev.\ D {\bf 66}, 071102 (2002)
[hep-ex/0208025].
%%CITATION = HEP-EX 0208025;%%

%\cite{Abe:2002np}
\bibitem{Abe:2002np}
K.~Abe {\it et al.}, BELLE Collaboration,
%``Study of Time-Dependent CP-Violating Asymmetries in b->sq q Decays,''
hep-ex/0212062.
%%CITATION = HEP-EX 0212062;%%

%\cite{Aubert:2003bq}
\bibitem{Aubert:2003bq}
B.~Aubert {\it et al.}, BABAR Collaboration,
%``Measurements of CP-violating asymmetries and branching fractions in B  meson decays to eta' K,''
hep-ex/0303046.
%%CITATION = HEP-EX 0303046;%%

%\cite{Aubert:2002nx}
\bibitem{Aubert:2002nx}
B.~Aubert {\it et al.}, BABAR Collaboration,
%``Measurement of sin(2beta) in B0 $\to$ Phi K0(S),''
hep-ex/0207070.
%%CITATION = HEP-EX 0207070;%%

%\cite{Gronau:1989ia}
\bibitem{Gronau:1989ia}
M.~Gronau,
%``CP Violation In Neutral B Decays To CP Eigenstates,''
Phys.\ Rev.\ Lett.\  {\bf 63}, 1451 (1989).
%%CITATION = PRLTA,63,1451;%%

%\cite{Grossman:2002bu}
\bibitem{Grossman:2002bu}
Y.~Grossman, A.~L.~Kagan and Z.~Ligeti,
%``Can the CP asymmetries in B $\to$ psi K(S) and B $\to$ psi K(L) differ?,''
Phys.\ Lett.\ B {\bf 538} (2002) 327
[hep-ph/0204212].
%%CITATION = HEP-PH 0204212;%%

%\cite{Savage:ub}
\bibitem{Savage:ub}
M.~J.~Savage and M.~B.~Wise,
%``SU(3) Predictions For Nonleptonic B Meson Decays,''
Phys.\ Rev.\ D {\bf 39}, 3346 (1989)
[Erratum-ibid.\ D {\bf 40}, 3127 (1989)].
%%CITATION = PHRVA,D39,3346;%%

%\cite{Grinstein:1996us}
\bibitem{Grinstein:1996us}
B.~Grinstein and R.~F.~Lebed,
%``SU(3) Decomposition of Two-Body B Decay Amplitudes,''
Phys.\ Rev.\ D {\bf 53}, 6344 (1996)
[hep-ph/9602218].
%%CITATION = HEP-PH 9602218;%%

%\cite{Grossman:1997gr}
\bibitem{Grossman:1997gr}
Y.~Grossman, G.~Isidori and M.~P.~Worah,
%``CP asymmetry in B/d $\to$ Phi K(S): Standard model pollution,''
Phys.\ Rev.\ D {\bf 58}, 057504 (1998)
[hep-ph/9708305].
%%CITATION = HEP-PH 9708305;%%

%\cite{Dighe:1997wj}
\bibitem{Dighe:1997wj}
A.~S.~Dighe, M.~Gronau and J.~L.~Rosner,
%``B decays to charmless V P final states,''
Phys.\ Rev.\ D {\bf 57}, 1783 (1998)
[hep-ph/9709223];
%%CITATION = HEP-PH 9709223;%%
%\cite{Dighe:1995bm}
%\bibitem{Dighe:1995bm}
A.~S.~Dighe,
%``Determination of CKM phases through rigid polygons of flavor SU(3) amplitudes,''
Phys.\ Rev.\ D {\bf 54}, 2067 (1996)
[hep-ph/9509287].
%%CITATION = HEP-PH 9509287;%%

%\cite{Deshpande:2000jp}
\bibitem{Deshpande:2000jp}
N.~G.~Deshpande, X.~G.~He and J.~Q.~Shi,
%``SU(3) flavor symmetry and CP violating rate differences for charmless  B $\to$ P V decays,''
Phys.\ Rev.\ D {\bf 62}, 034018 (2000)
[hep-ph/0002260].
%%CITATION = HEP-PH 0002260;%%

%\cite{London:1997zk}
\bibitem{London:1997zk}
D.~London and A.~Soni,
%``Measuring the CP angle beta in hadronic b $\to$ s penguin decays,''
Phys.\ Lett.\ B {\bf 407}, 61 (1997)
[hep-ph/9704277].
%%CITATION = HEP-PH 9704277;%%

%\cite{Hagiwara:fs}
\bibitem{Hagiwara:fs}
K.~Hagiwara {\it et al.}, Particle Data Group,
%``Review Of Particle Physics,''
Phys.\ Rev.\ D {\bf 66}, 010001 (2002).
%%CITATION = PHRVA,D66,010001;%%

%\cite{Abe:2002ms}
\bibitem{Abe:2002ms}
K.~Abe {\it et al.}, BELLE Collaboration,
%``Study of charmless B decays to three-kaon final states,''
hep-ex/0208030.
%%CITATION = HEP-EX 0208030;%%

%\cite{Gronau:2003ep}
\bibitem{Gronau:2003ep}
M.~Gronau and J.~L.~Rosner,
%``I-spin and U-spin in B $\to$ K K anti-K,''
hep-ph/0304178.
%%CITATION = HEP-PH 0304178;%%

%\cite{Aubert:2003xz}
\bibitem{Aubert:2003xz}
B.~Aubert {\it et al.}, BABAR Collaboration,
%``Measurements of the branching fractions and charge asymmetries of  charmless three-body charged B decays,''
hep-ex/0304006.
%%CITATION = HEP-EX 0304006;%%

%\cite{Falk:2001hx}
\bibitem{Falk:2001hx}
A.~F.~Falk, Y.~Grossman, Z.~Ligeti and A.~A.~Petrov,
%``SU(3) breaking and D0 - anti-D0 mixing,''
Phys.\ Rev.\ D {\bf 65}, 054034 (2002)
[hep-ph/0110317].
%%CITATION = HEP-PH 0110317;%%


\end{thebibliography}
\end{document}